\documentclass[conference]{IEEEtran}
\usepackage{fancyhdr}
\usepackage{textcomp} 

\usepackage[T1]{fontenc}
\usepackage[utf8]{inputenc}
\usepackage{listings}
\usepackage{graphicx}
\usepackage{amsmath,amssymb,amsfonts}
\usepackage{xcolor}
\usepackage{placeins}
\usepackage{url}
\usepackage{hyperref}
\usepackage{adjustbox}
\usepackage{float}
\usepackage{booktabs}
\usepackage{flafter}
\usepackage{stfloats}
\begin{document}
	\title{Bridging In-Situ Visualization to Game Engines : From ParaView Catalyst via ANARI to Unreal Engine}
	
	\author{
		\IEEEauthorblockN{Thomas George}
		\IEEEauthorblockA{
			Forschungszentrum J\"ulich GmbH\\
			J\"ulich Supercomputing Centre\\
			J\"ulich, Germany\\
			t.george@fz-juelich.de}
		\and
		\IEEEauthorblockN{Severin A. T. Klapproth}
		\IEEEauthorblockA{
			Paul Scherrer Institute (PSI)\\
			Center for Scientific Computing, Theory and Data\\
			Villigen, Switzerland\\
			severin.klapproth@psi.ch}
		\and
		\IEEEauthorblockN{Jens Henrik G\"obbert}
		\IEEEauthorblockA{
			Forschungszentrum J\"ulich GmbH\\
			J\"ulich Supercomputing Centre\\
			J\"ulich, Germany\\
			j.goebbert@fz-juelich.de}
		\and
		\IEEEauthorblockN{Jonathan Windgassen}
		\IEEEauthorblockA{
			Forschungszentrum J\"ulich GmbH\\
			J\"ulich Supercomputing Centre\\
			J\"ulich, Germany\\
			j.windgassen@fz-juelich.de}
		\and
		\IEEEauthorblockN{Andreas Adelmann}
		\IEEEauthorblockA{
			Paul Scherrer Institute (PSI)\\
			Center for Scientific Computing, Theory and Data\\
			Villigen, Switzerland\\
			andreas.adelmann@psi.ch}
		\and
		\IEEEauthorblockN{Sriramkrishnan Muralikrishnan}
		\IEEEauthorblockA{
			Forschungszentrum J\"ulich GmbH\\
			J\"ulich Supercomputing Centre\\
			J\"ulich, Germany\\
			s.muralikrishnan@fz-juelich.de}
	}
	
	\maketitle
	\thispagestyle{fancy}
	\lhead{}
	\rhead{}
	\chead{}
	\lfoot{}
	\rfoot{}
	\cfoot{\parbox[t]{\textwidth}{\scriptsize\centering
	SC26 Workshops, November 15--20, 2026, Chicago, Illinois, USA\\
	979-8-3195-1221-5/26/\$31.00~\copyright~2026 IEEE. Personal use of this material is
	permitted. Permission from IEEE must be obtained for all other uses, in any current or
	future media, including reprinting/republishing this material for advertising or
	promotional purposes, creating new collective works, for resale or redistribution to
	servers or lists, or reuse of any copyrighted component of this work in other works.
	Accepted for publication in the Proceedings of the SC26 International Conference for
	High Performance Computing, Networking, Storage, and Analysis (SC26).}}
	\renewcommand{\headrulewidth}{0pt}
	\renewcommand{\footrulewidth}{0pt}

	\begin{abstract}
		In-situ high-performance computing (HPC) simulations produce massive datasets that remain difficult to stream into game engines for interactive visualization. We present an in-situ, disk-free pipeline that couples simulation data to Unreal Engine 5 through ParaView's ANARI rendering API. A modified ANARI-USD device captures Universal Scene Description (USD) scene graphs in memory at the ANARI layer and streams the engine-agnostic USD format through an embedded ZeroMQ broker to visualization clients, where JuSync, middleware with a companion Unreal Engine 5 plugin, parses the geometry as it arrives and routes it to engine rendering components at runtime, producing scenes that are camera-navigable once received. The decoupled design keeps simulation execution independent of visualization load, and clients may connect or disconnect at any time. The pipeline is demonstrated on the JURECA-DC supercomputer at Forschungszentrum J\"ulich, streaming Free Electron Laser (FEL) simulation data with point clouds of tens of millions of particles.
	\end{abstract}
	
	\begin{IEEEkeywords}
		In-Situ, ANARI, Scientific visualization, Unreal Engine, Real-time rendering, HPC, Streaming
	\end{IEEEkeywords}
	
	\section{Introduction}
	
	\begin{figure*}[t]
		\centering
		\includegraphics[width=0.8\textwidth]{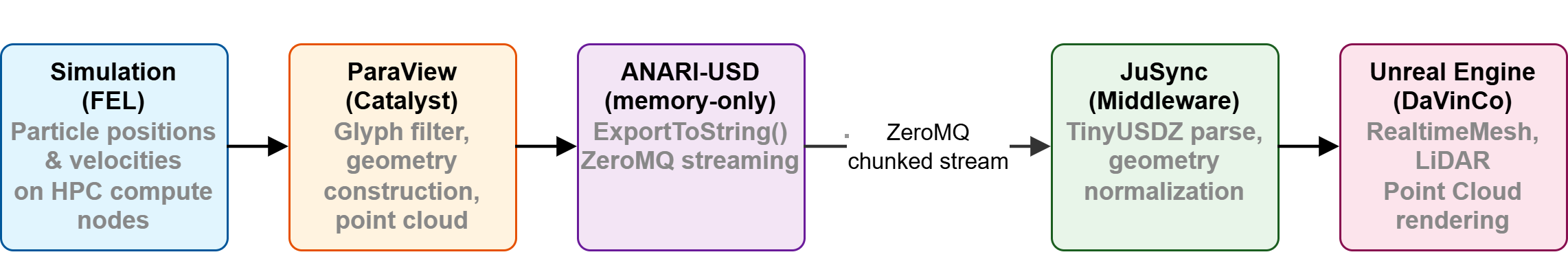}
		\caption{Overview of the in-situ streaming pipeline. Simulation data flows through
			ParaView and a modified ANARI-USD device that embeds ZeroMQ streaming, to the
			JuSync middleware with its companion UE5 plugin, delivering interactive visualization
			in Unreal Engine. Arrows indicate the direction of data flow.}
		\label{fig:pipeline}
	\end{figure*}
	
	ParaView, paired with its Catalyst framework, is one of the standard tools for in-situ analysis on HPC clusters, keeping data in memory without writing to disk. ParaView targets scientific analysis, whereas interactive presentation is the domain of a complementary tool --- a modern game engine. Rendering simulation data inside realistic, contextually accurate environments --- for example, a reconstructed laboratory facility --- has the potential to increase scientific insight by letting researchers explore their results as they would look in the real world. Game engines such as Unreal Engine 5 are built for exactly these capabilities: immersive virtual reality (VR) navigation, realistic lighting, custom annotations, and multi-user collaboration. However, Unreal Engine's built-in USD support and Datasmith~\cite{datasmith2022} import data from files on disk, not from live streaming in memory.
	
	This work bridges that gap by streaming USD scene graphs directly from ParaView to visualization clients, disk-free, yielding scenes that are interactable on the client once the payload is received. Because JuSync operates at the USD level, the streamed data could in principle feed rendering engines other than Unreal Engine, though only the Unreal Engine 5 client is demonstrated in this work. 
	
	As shown in Figure~\ref{fig:pipeline}, the pipeline moves data from the simulation on the HPC cluster through ParaView and a memory-only ANARI-USD device, across the network via ZeroMQ, and into the JuSync middleware before finally rendering in Unreal Engine. The dashed line separates the cluster-side processing from the client-side visualization.
	
	Following the pipeline in Figure~\ref{fig:pipeline}, three gaps appear along the data path. First, on the simulation side, the simulation produces particle data that ParaView processes in-situ and ANARI-USD converts into USD scene graphs in memory. However, those graphs stay on the cluster and are not sent over a network. Second, on the client side, even if the USD payload is streamable, Unreal Engine cannot consume it directly at runtime. The scene graph must be parsed, converted to triangulated geometry, and routed to engine-specific rendering components for which no standard tool exists. Finally, the transport must keep the simulation running independently of network conditions or client speed.
	
	These challenges are addressed with two components, illustrated in Figure~\ref{fig:pipeline}. The ANARI-USD device is extended with a memory tracking layer and an embedded ZeroMQ transport that streams USD scene graphs directly from RAM; the broker is designed to support multiple visualization clients. JuSync~\cite{jusync2026} is presented as a standalone C++ middleware that receives the streamed USD over ZeroMQ and parses it into runtime geometry. Together, these form a complete disk-free path from ParaView to Unreal Engine. The additions are confined to a modified ANARI-USD device and the JuSync client plugin; the simulation, ParaView, and the Unreal Engine core itself require no code changes.
	
	Our contributions are:
	
	\begin{itemize}
		\item A memory-only ANARI-USD serialization path that streams standardized USD scene graphs from ParaView over ZeroMQ, enabling a disk-free path to a game engine that is engine-agnostic at the USD boundary.
		\item JuSync, an engine-agnostic C++ middleware with a companion UE5 plugin, bridging the gap between streamed USD scene graphs and runtime Unreal Engine geometry.
	\end{itemize}
	
	For results and evaluation:
	\begin{itemize}
		\item An end-to-end evaluation of the pipeline on the JURECA-DC supercomputer using a Free Electron Laser (FEL) simulation with point clouds of tens of millions of particles. The pipeline itself is not limited to FEL and works with any ParaView-compatible input.
	\end{itemize}
	
	\section{State of the Art}
	\label{sec:stateoftheart}
	
	\subsection{ParaView-to-External-Renderer Pipelines}
	Several recent pipelines share the goal of bridging ParaView's scientific data processing with a game engine. These pipelines face a common set of challenges: they must transfer scene data produced by ParaView's visualization pipeline to a rendering environment that operates on a different runtime, often on a different machine, without introducing prohibitive latency.
	
	InsitUE~\cite{insitue2024} addresses this with a Catalyst2-compatible hybrid workflow that streams simulation results directly into Unreal Engine. It supports both in-situ and in-transit operation, allows bidirectional steering of visualization parameters, and uses a custom TCP/UDP socket communication protocol with TCP for the control plane and UDP for the data plane. InsitUE couples a Catalyst2/Conduit path to a pair of components --- its Catalyst implementation and its Unreal plugin --- and additionally offers bidirectional simulation steering, which we do not provide. The two systems differ chiefly in the interchange format: InsitUE transfers geometry as Conduit mesh blueprints, whereas our pipeline emits USD scene graphs in memory at the ANARI layer, a format that existing USD-capable renderers can ingest. ZeroMQ then delivers these graphs to a client that consumes USD alone, though only the Unreal Engine client is demonstrated here. The sort-last in-transit approach by Nealey et al.~\cite{nealey2022} uses SENSEI~\cite{sensei2016} to move data to staging nodes, where ParaView Catalyst extracts PLY mesh geometry that is forwarded to Unreal Engine over WebSockets, demonstrating near-real-time visualization of distributed simulation data. Kitware has demonstrated VTK rendering within Unity~\cite{kitware2020unity} using a direct VTK-to-Unity bridge. A ParaView-to-Unity VR pipeline has been developed for cosmological particle data, extracting VTK geometry from ParaView and importing it into Unity for immersive cluster analysis~\cite{adeniji2024}. Outside the game-engine domain, ParaView has been used with Blender to generate augmented reality (AR) models from finite element simulation data~\cite{mathur2022}, where ParaView extracts surface geometry and Blender handles rendering and export. 	The NVIDIA Omniverse ParaView Connector~\cite{nvidiaanari2023} takes a different approach: it uses the ANARI-USD device to write USD stages to shared storage, which Omniverse then watches and ingests, allowing Omniverse's rendering engine to access ParaView-generated scenes without tight coupling.

Across these pipelines, scene data reaches the engine in a representation tied to the data path---VTK or Conduit geometry mapped by a dedicated bridge, PLY meshes over WebSockets, or USD stages written to shared storage---each a reasonable fit for its own goal. We instead standardize the interchange on USD captured in memory at ParaView's ANARI layer, removing the shared-filesystem round-trip and confining engine-specific work to a thin client plugin. We chose USD for practical, not superiority, reasons: it describes a full scene (hierarchy, transforms, materials, and metadata) rather than geometry alone; it is what the ANARI-USD device already emits, so no producer-side conversion is needed; and its open-source support across renderers underpins the engine-agnostic boundary. Beyond reaching a game engine, a standard scene format also keeps the payload inspectable and reusable, which aids debugging and reproducibility. Decoupling producers from consumers is the design aim elaborated in Section~\ref{sec:architecture}. We do not claim to remove every constraint of the surveyed systems: like them, we demonstrate a single renderer (Unreal Engine); unlike InsitUE, we provide no simulation steering; and in our evaluation, forwarding latency currently dominates end-to-end time (Section~\ref{sec:evaluation}).

	\subsection{ANARI as a Rendering Abstraction}
	The Khronos ANARI specification~\cite{anari2023} defines a device-based rendering abstraction that decouples applications from specific rendering backends such as OSPRay and NVIDIA VisRTX. Applications issue rendering commands through the ANARI API, and any conformant backend device translates them into its native rendering operations. This design allows visualization applications like ParaView to switch between external rendering backends without code changes.
	
	The ANARI-USD device (NVIDIA's AnariUsdDevice)~\cite{nvidiaanari2023} can serve as such an external ``rendering'' backend. Even though it does not render any image, it produces USD scene descriptions from the ANARI rendering commands. It receives ANARI scene graphs from ParaView's rendering pipeline and writes the resulting USD stages to shared storage, where applications such as NVIDIA Omniverse can ingest them for interactive visualization. The device handles geometry conversion, material assignment, and scene graph construction automatically, making the ParaView-to-USD path transparent to the user.
	
	This work builds on the same ANARI-USD device but replaces the disk-based write path with memory-only serialization and ZeroMQ streaming. The ANARI scene graph is captured from an in-memory serialization path, stored in a per-rank memory buffer (one buffer per MPI process), and streamed over the network. This enables scene transfer to external rendering engines without shared filesystem access, while retaining all the benefits of the ANARI abstraction layer.
	
	\subsection{In-Situ Data Movement}
	Moving simulation data to visualization consumers with minimal overhead is a core challenge in in-situ workflows. ParaView Catalyst Live~\cite{catalystlive2023} provides real-time streaming of in-memory VTK data to a live ParaView GUI connection, enabling interactive exploration during simulation execution. Catalyst-ADIOS2~\cite{catalyst-adios2023} integrates ADIOS2~\cite{adios22020} with ParaView Catalyst for in-transit streaming, supporting data reduction and transfer over high-speed interconnects. It uses ADIOS2's staging engines to move data between simulation and visualization nodes without writing intermediate files. Ascent~\cite{ascent2022} takes a flyweight approach, providing a lightweight in-situ library that minimizes the simulation-side integration burden while supporting multi-stage analysis pipelines. Inshimtu~\cite{inshimtu2023} provides a lightweight shim between simulation codes and ParaView, adapting simulation data structures for visualization consumption without requiring the simulation to link against ParaView's libraries. It handles data format conversion and memory management, allowing ParaView to connect to simulations that would otherwise require custom instrumentation. Serviz~\cite{serviz2022} proposes a shared in-situ visualization service that runs alongside the simulation and serves visualization data to multiple consumers on demand. By decoupling producers from consumers, Serviz allows different visualization tools to connect to the same simulation independently.
	
	This project uses ParaView Catalyst for data movement from the simulation to the ParaView side. On the client side of the pipeline presented here, JuSync receives streamed USD payloads from a modified ANARI-USD device, parses them with TinyUSDZ, and routes the resulting geometry to rendering components, including Epic's LiDAR Point Cloud plugin~\cite{epiclidar2023} and the RealtimeMesh plugin~\cite{realtimemesh2023}, which support runtime updates from streaming data sources.
	
	
	
	
	\section{Free Electron Laser Undulator Simulation}
	\label{sec:FEL}
	
	As the in-situ data source for our pipeline, we use a Free Electron Laser (FEL) undulator simulation originally developed as a mini-application in the IPPL (Independent Parallel Particle Layer) framework~\cite{muralikrishnan2024scaling,matthias_frey_2024_10878166}. IPPL is a massively parallel high-performance computing library providing parallel data structures and operator primitives for performance-portable particle-mesh methods, with primary applications in beam/plasma physics and cosmology. The FEL simulation models the propagation of an electron beam through a periodic undulator magnet array, generating coherent X-ray radiation as beam particles oscillate and interact with their self-consistent electromagnetic fields. The code supports configurable particle and mesh resolutions, producing datasets with particle positions and velocity vectors for up to billions of particles, as well as electromagnetic field data on uniform refined grids. In the near future, this will allow simulations to abandon the approximate modelling via macroparticles and instead simulate real electrons. In a recent update, IPPL was instrumented with the Catalyst API to support direct in-situ export of its core data structures~\cite{Klapproth2026} enabling any IPPL-based simulation to serve as a data source for our visualization pipeline with minimal integration effort.
	
	The FEL simulation was selected as the demonstration case because it represents a scientifically significant and computationally challenging regime, rooted in the specific nature of its real-world experimental counterpart. There is considerable interest in the particle accelerator community in better understanding electron beam dynamics, in particular the coupled processes of intrabeam scattering and microbunching. Current standard simulation codes encounter difficulties in capturing these phenomena at the accuracy required~\cite{prat2022}: at the Swiss Free Electron Laser, intrabeam scattering has been observed to degrade beam quality, suppressing microbunch formation and reducing achievable photon beam brightness. This effect was not predicted by prior simulations and motivates the need for advanced simulation models and analysis, capable of studying these processes in finer detail. Accurately modeling Coulomb collisions between electrons, which lie at the root of intrabeam scattering, requires abandoning the macroparticle approximation~\cite{lucas2026} in favour of tracking real electrons. At the scale of a realistic accelerator, this pushes the total particle count to order $O(10^9)$, producing datasets too large to reasonably write and store to disk. This makes the FEL simulation a prime example of a high-performance scientific code unsuited to rely on traditional post-processing workflows and instead benefitting directly from modern in-situ analysis --- enabling detailed 3D visualization that captures in full the phenomena advanced models are designed to resolve.
	
	The FEL undulator simulation is therefore integrated into the pipeline via the IPPL--Catalyst interface, coupling the simulation directly to ParaView without intermediate file I/O. At each user-defined timestep, particle positions, velocity vectors, and electromagnetic field data are passed in-situ to ParaView~\cite{catalyst2015} through Catalyst, which constructs visual representations and feeds the resulting scene graph to the modified ANARI-USD device.

	\begin{figure*}[t]
		\centering
		\includegraphics[width=0.75\linewidth]{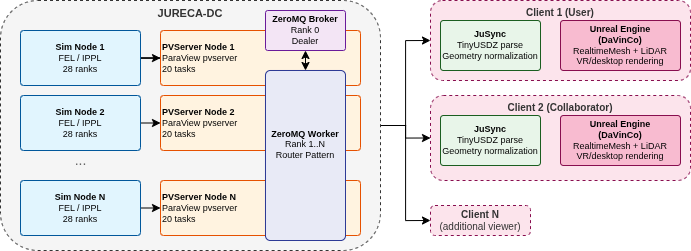}
		\caption{HPC Network Topology showing the JURECA node allocation and ZeroMQ routing via an SSH tunnel to the visualization client.}
		\label{fig:hpc_topology}
	\end{figure*}

	\section{Pipeline Architecture}
	\label{sec:architecture}

	\subsection{Simulation Layer}

	The pipeline works with any ParaView-compatible input. This work demonstrates it using the FEL simulation described in Section~\ref{sec:FEL}, which produces particle positions and velocity vectors that are made available to ParaView through an in-situ workflow using ParaView Catalyst. The FEL simulation can push millions of particles per timestep, limited only by available resources.
	
	The simulation and ParaView run concurrently on dedicated node pools, with the simulation generating particle data while ParaView constructs and updates the visualization pipeline and the modified ANARI-USD device produces in-memory USD stages (deployment, as shown in Figure~\ref{fig:hpc_topology}). The specific node counts and rank layout used for evaluation are given in Section~\ref{sec:evaluation}.
	
	\subsection{ParaView and ANARI Integration}
	ParaView performs visualization-oriented data processing. 	In the evaluation setup, the FEL particle data is processed as point clouds and rendered directly as points; ParaView's Glyph filter would convert particle positions into sphere glyphs only if triangulated mesh geometry were required, a mode not exercised in this evaluation. To decouple visualization construction from the rendering backend, ParaView
	interfaces with ANARI, which provides a device-based rendering abstraction that
	allows different rendering devices to be used, such as VisRTX, OSPRay, or ANARI-USD.
	The ANARI-USD device implements this interface by producing USD scene
	descriptions from ANARI rendering commands rather than generating rasterized
	images.
	Importantly, ParaView itself remains unchanged; all pipeline-specific
	modifications are confined to the ANARI-USD device.
	
	\subsection{USD Memory Streaming}
	The ANARI-USD device for ParaView supports MPI where each rank produces USD scene graphs independently. By default, each committed stage is serialized to disk, adding filesystem I/O to the per-timestep path at multi-node scale. The modification introduces a tracking layer on the device's existing in-memory serialization path: each committed USD stage is captured into a per-rank memory store as a serialized buffer, with an XXH3-128 hash computed on each payload for fast change detection. The disk write step is bypassed entirely.
	
\begin{figure}[!t]
		\centering
		\includegraphics[width=\linewidth,height=0.6\textheight,keepaspectratio]{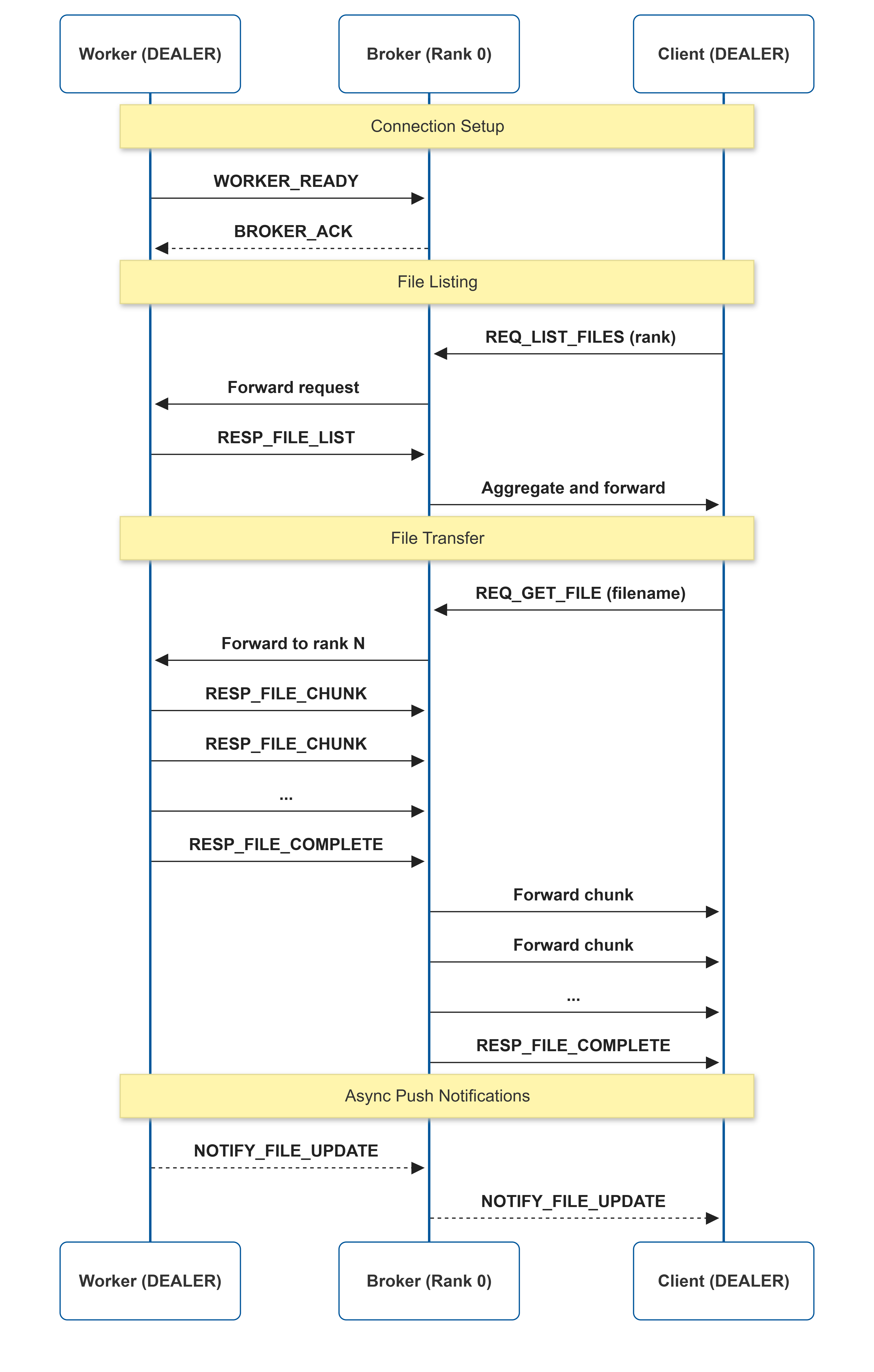}
		\caption{Sequence diagram of the ZeroMQ messaging protocol, illustrating the Request/Response and asynchronous Push notifications between the Broker and the client.}
		\label{fig:zmq_sequence}
	\end{figure}

	On top of this, a ZeroMQ transport layer streams the in-memory buffers across the interconnect to visualization clients (protocol summarized in Figure~\ref{fig:zmq_sequence}). On each worker rank, a \texttt{DEALER} socket connects to a broker on rank~0. The broker runs two \texttt{ROUTER} sockets: one for worker connections and one for client connections. The broker aggregates file metadata from all workers into a unified reference table, while the actual USD payload data remains distributed in the per-rank memory stores. When a client requests a file, the broker routes the request to the appropriate worker rank, which streams the data back through the broker in configurable chunks (default $4$\,MB). Chunking keeps per-transfer memory bounded and predictable on the client while adding negligible framing overhead, and the granular flow lets parsing and streaming overlap. The chunk size is set primarily as a fault-tolerance granularity: because the link crosses an external SSH tunnel, a drop mid-transfer costs at most the remainder of the in-flight dataset, and a reconnected client resumes from the broker's dataset table with hash-based dedup skipping datasets already received.
	
	This memory-only approach removes any dependence on the parallel filesystem from the data path: ANARI rendering commands are converted by the USD bridge and serialized into per-rank memory stores, and shared files are relayed through the broker to clients without ever touching disk. The broker itself holds only the aggregated file list; the USD payload data remains distributed in the per-rank memory stores until a client pulls it.
	
	\subsection{JuSync Middleware}
	\label{sec:implementation}
	
	\begin{figure}[!t]
		\centering
		\includegraphics[width=\linewidth,height=0.5\textheight,keepaspectratio]{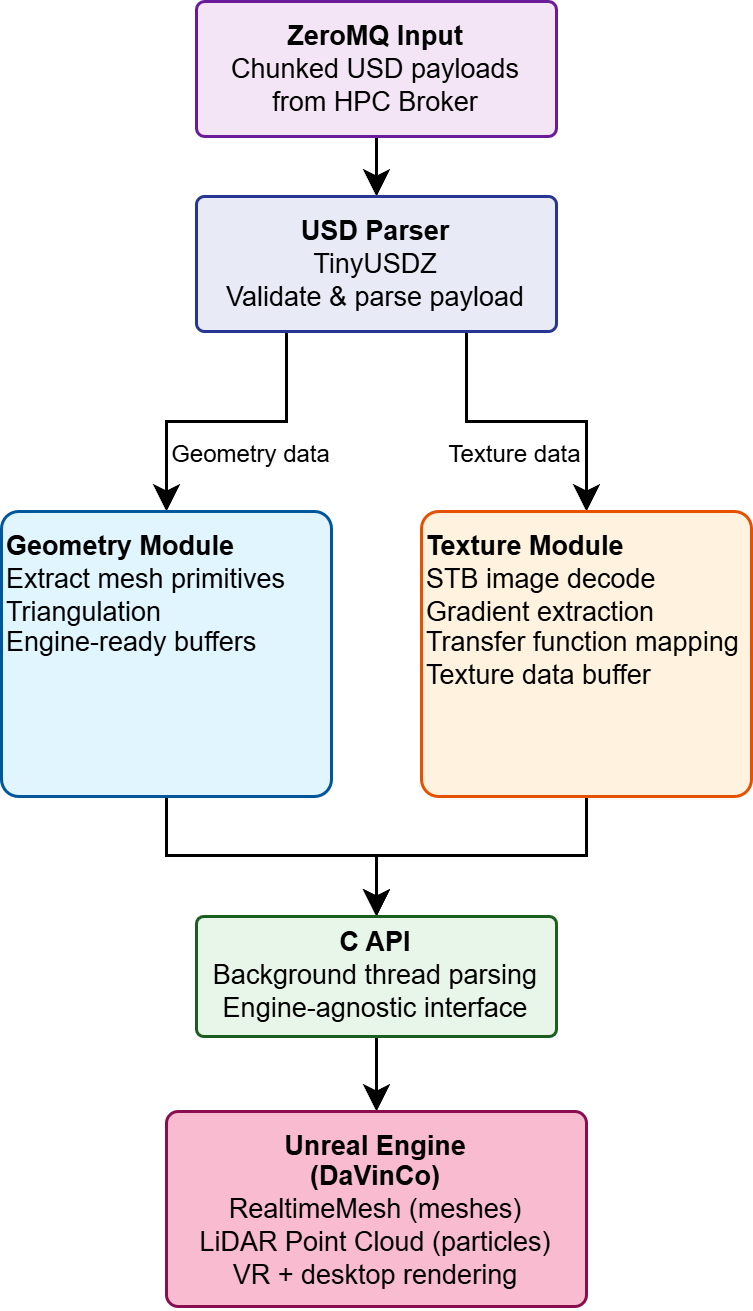}
		\caption{Data Flow from the ZeroMQ Router through the JuSync Middleware, showing the separated parsing pipelines for USD geometry and texture data.}
		\label{fig:zmq_middleware}
	\end{figure}
	
	JuSync~\cite{jusync2026} is a standalone C++ middleware that receives USD payloads from the modified ANARI-USD device via ZeroMQ streaming (data flow in Figure~\ref{fig:zmq_middleware}). Each payload is parsed with TinyUSDZ~\cite{tinyusdz2026} for lightweight, dependency-free USD parsing. The geometry module extracts mesh primitives into an internal representation, converting them into engine-ready buffers with required coordinate transforms (e.g., right-handed to left-handed system conversion) and triangulation. Texture data are decoded via the STB~\cite{stb2026} library, with gradients extracted for transfer-function mapping. The middleware exposes a C API intended for use by other rendering engines, with background threads handling parse operations asynchronously.
	
	\subsection{Decoupled Architecture}
	The modifications are confined to the ANARI-USD device (tracking layer and ZeroMQ streaming) and the JuSync middleware; the simulation, ParaView, and Unreal Engine themselves require no code changes. Deployment nevertheless requires building and linking the modified ANARI-USD device, running the ZeroMQ broker, and installing the JuSync UE5 plugin (Appendix), rather than the stock ParaView/Unreal stack alone. Because JuSync sits at the USD level, any simulation that feeds into ParaView can be streamed to Unreal Engine without code modifications to the simulation or ParaView. The architecture provides decoupling along four independent axes. \textit{(i) Producer/consumer decoupling:} the simulation and ParaView continue running regardless of whether a client is connected, and clients can join or leave at any time. \textit{(ii) HPC/client decoupling:} the Unreal instance requires only TinyUSDZ for lightweight USD parsing rather than the full OpenUSD stack or ParaView's binary dependencies. \textit{(iii) Simulation/rendering decoupling:} ParaView handles scientific data processing and visualization construction, while Unreal Engine handles real-time rendering and interaction. \textit{(iv) Simulation/visualization decoupling:} the asynchronous ZeroMQ transport is designed so that USD payload transmission does not block the simulation, because the broker buffers messages independently of client timing.
	
	\subsection{Unreal Engine Plugin Integration}
	A companion UE5 runtime plugin~\cite{jusync2026} incorporates the JuSync middleware. On startup, it initializes the C middleware library, connects to the ZeroMQ broker, and receives USD payloads through the C API. Received data are dispatched to Unreal Engine's rendering components on the game thread.
	
	Point cloud data from particle simulations are routed through Epic's LiDAR Point Cloud plugin~\cite{epiclidar2023}, which provides dynamic level-of-detail (LOD) scaling, global point budget control, and runtime insertion capabilities for large-scale particle datasets. Mesh data are streamed into \texttt{URealtimeMeshComponent} for interactive triangulated geometry. Both rendering paths update at runtime without intermediate file I/O. DaVinCo~\cite{davinco2022}, a collaborative multi-user VR/desktop visualization application developed by Forschungszentrum J\"ulich, provides the rendering environment driven by this plugin.
	
	\section{Results and Evaluation}
	\label{sec:evaluation}
	
	\subsection{Setup}
	Evaluation was conducted on the JURECA-DC system at Forschungszentrum J\"ulich (dc-cpu partition). A 60-node allocation exercised the full pipeline at a representative multi-node HPC configuration, within the dc-cpu partition's practical limits. 52 nodes ran the FEL simulation and eight ran \texttt{pvserver} for ParaView processing ($64$ MPI worker ranks). The 52/8 split is a chosen trade-off rather than a tight quota: more simulation nodes shorten the FEL particle calculation, while more \texttt{pvserver} nodes give ParaView more workers for serializing and streaming USD, smoothing delivery to the client; we made no attempt to tune for optimal utilization, as the goal was multi-node scale rather than peak throughput. The client machine, equipped with an Intel Core i9-14700HX and an NVIDIA GeForce RTX 5090, ran Unreal Engine and connected via SSH tunnel (\texttt{-L 5556:127.0.0.1:5556}) to the ZeroMQ broker on MPI rank 0. 	Measurements focus on the client-side arrival path: each dataset size was streamed through the full pipeline and timed across the dataset-listing, receive, and parse stages, with transfer throughput and host memory usage recorded alongside. Cluster-side ANARI-USD serialization and per-frame rendering times were not captured and are not reported. Unless stated otherwise, reported times refer to the client-side arrival path over the external SSH-tunnel link, where network delivery dominates the wall-clock budget. The pipeline operates independently of whether the simulation on the HPC runs on CPU or GPU: it consumes whatever data ParaView produces in-situ, and the same path applies to any simulation feeding into ParaView via Catalyst, regardless of compute backend. The pipeline was also exercised on standard ParaView example datasets (mesh and point-cloud sources); those runs are not reported here, as the evaluation focuses on the production FEL use case.
	
	\subsection{Visualization Results}
\begin{figure}[htbp]
		\centering
		\includegraphics[width=0.8\linewidth]{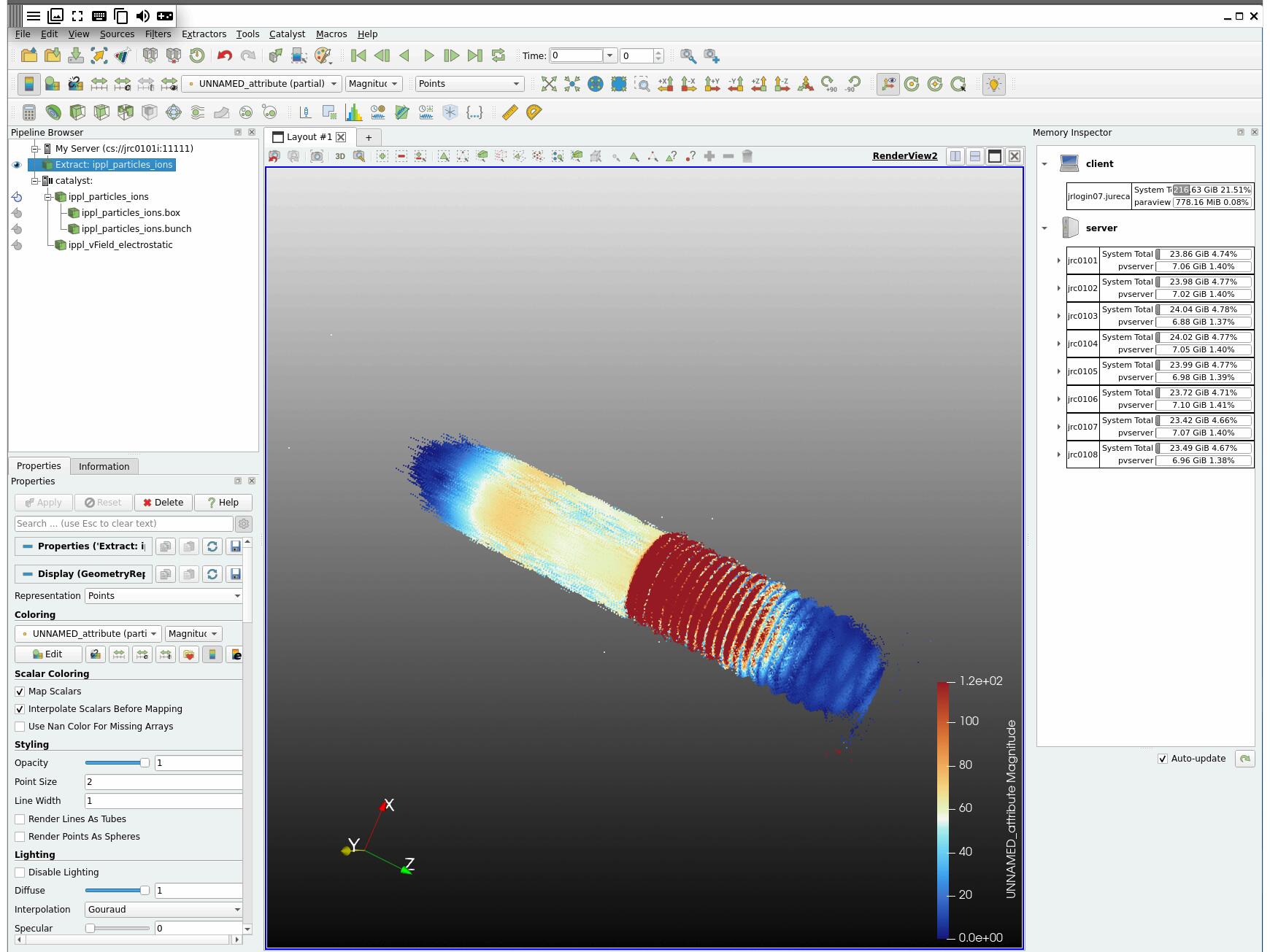}
		\caption{Full view of the FEL particle beam (40 million points) rendered in ParaView, colored by velocity magnitude.}
		\label{fig:fel_paraview_zoomout}
	\end{figure}

\begin{figure}[htbp]
		\centering
		\includegraphics[width=0.8\linewidth]{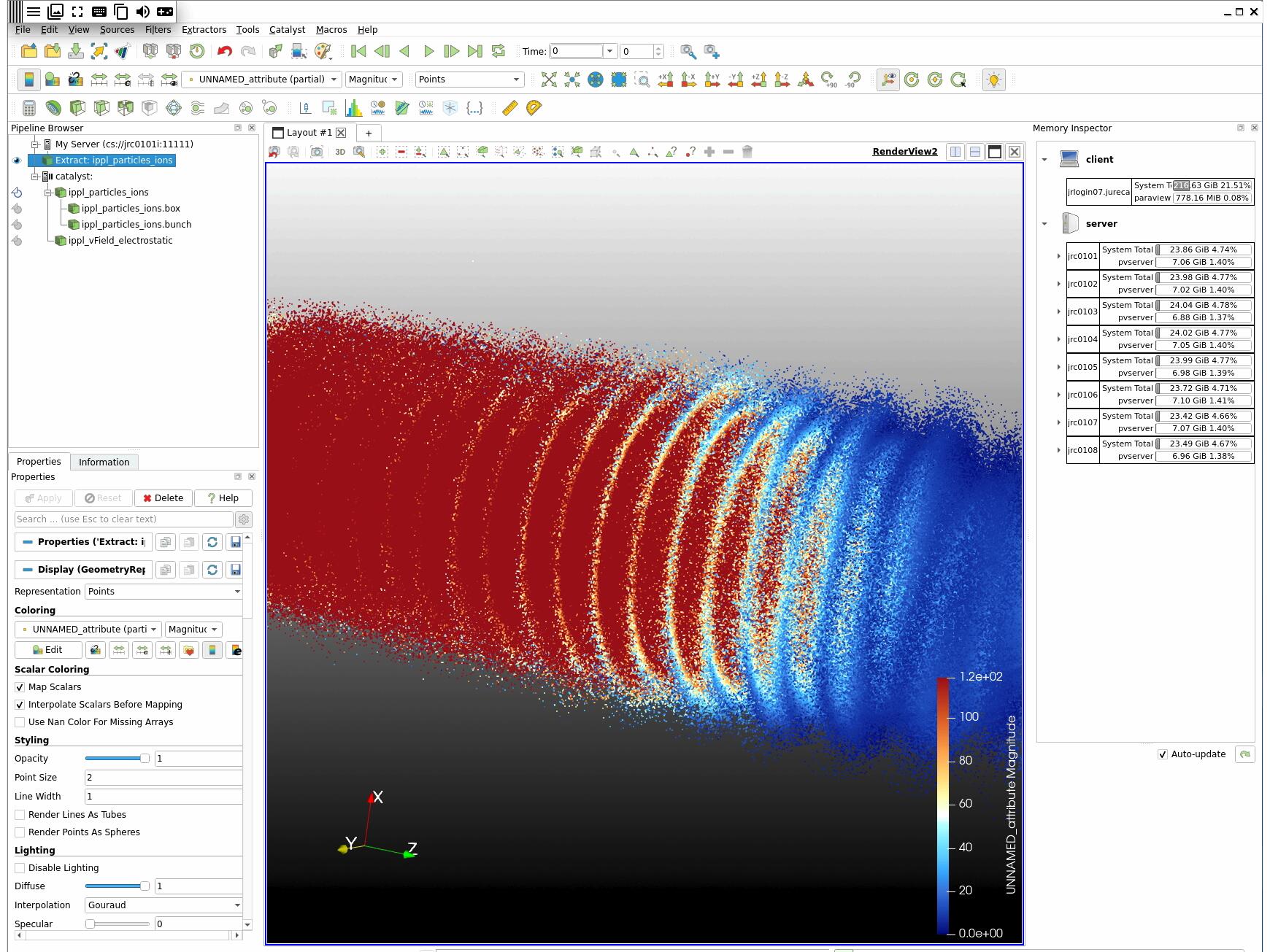}
		\caption{Zoomed-in view of the FEL particle beam in ParaView, revealing the microbunching structure and oscillation pattern under the undulator fields.}
		\label{fig:fel_paraview_zoomin}
	\end{figure}

\begin{figure}[htbp]
		\centering
		\includegraphics[width=0.8\linewidth]{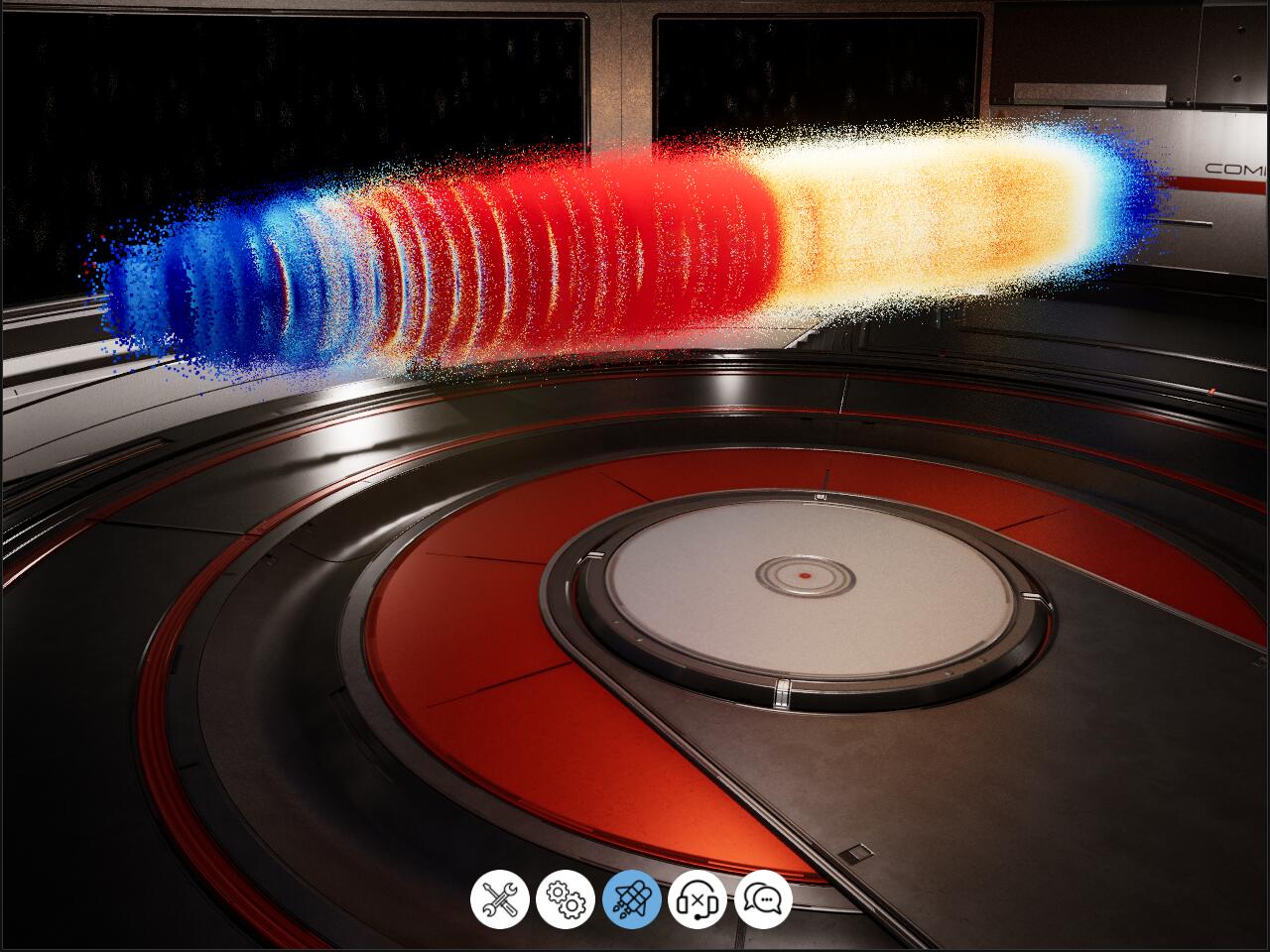}
		\caption{Full overview of the FEL point cloud as streamed through the pipeline and rendered in Unreal Engine via DaVinCo as a scene that is camera-navigable once received, using the LiDAR Point Cloud plugin.}
		\label{fig:fel_davinco_zoomout}
	\end{figure}

\begin{figure}[htbp]
		\centering
		\includegraphics[width=0.8\linewidth]{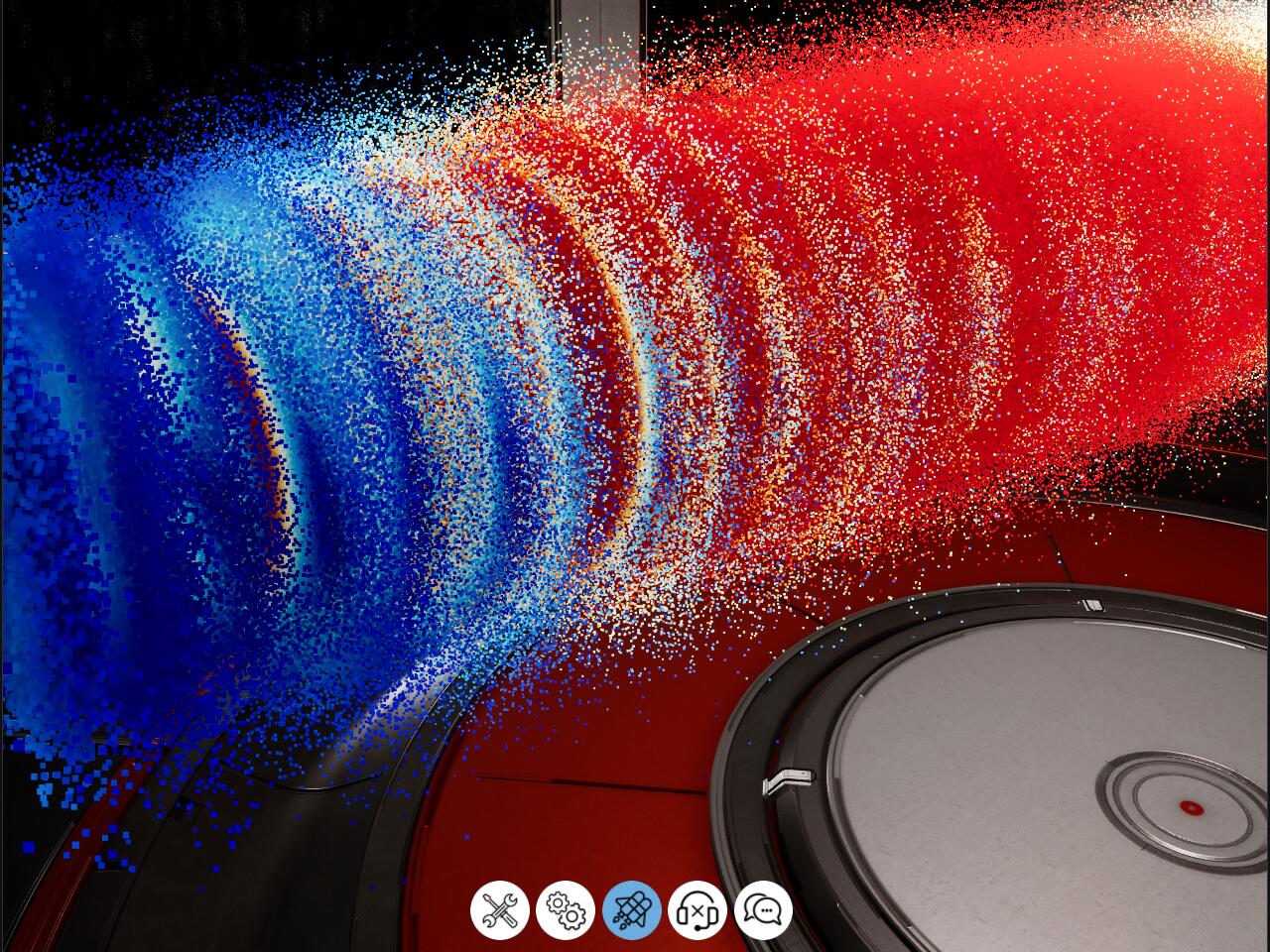}
		\caption{Close-up of the streamed FEL point cloud in DaVinCo, demonstrating interactive examination of microbunching structure with free camera navigation through the LiDAR Point Cloud plugin.}
		\label{fig:fel_davinco_zoomin}
	\end{figure}

\begin{figure}[htbp]
		\centering
		\includegraphics[width=0.8\linewidth]{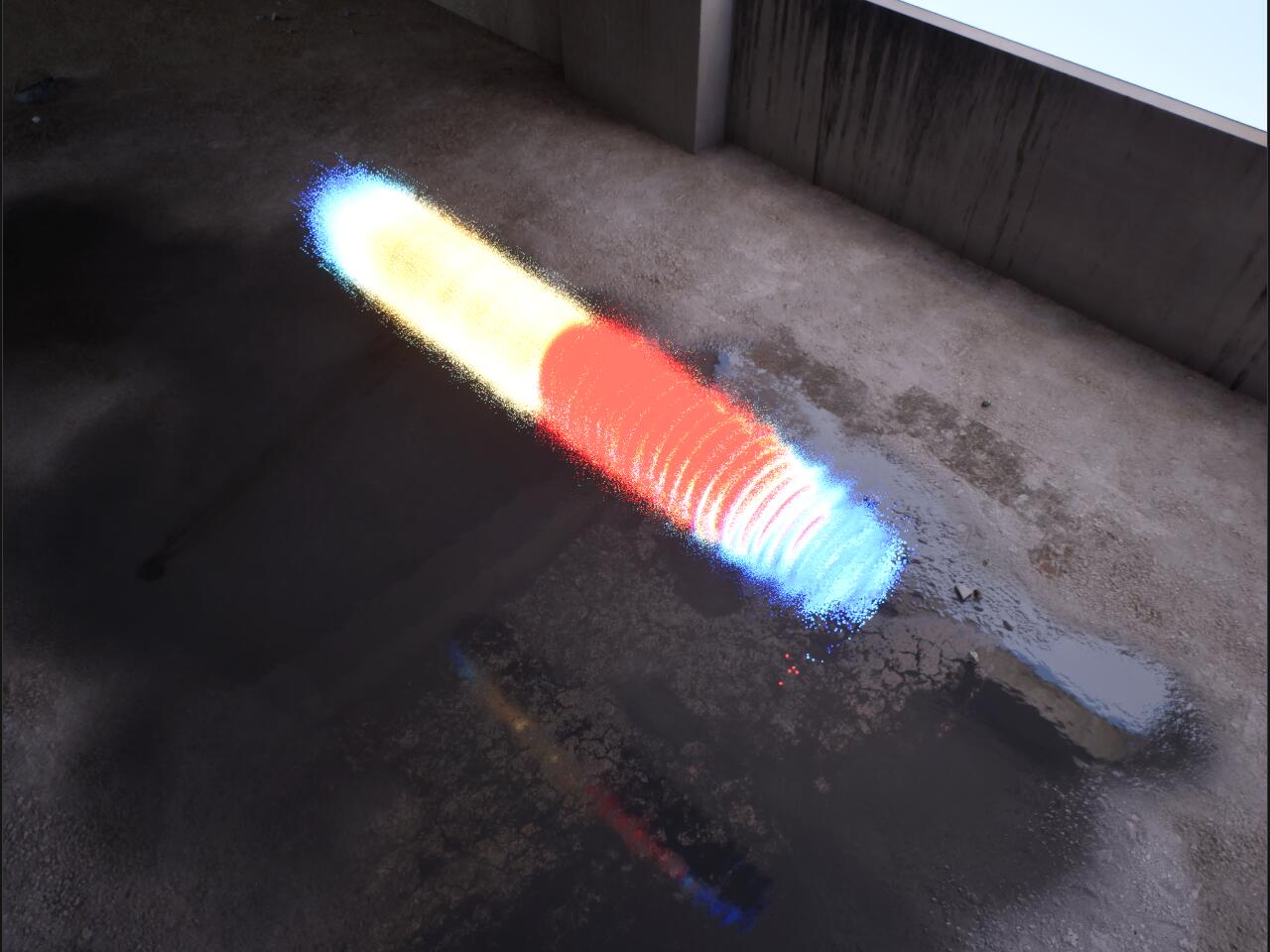}
		\caption{Alternate level in Unreal Engine demonstrating achievable visual fidelity with ray tracing, physically-based reflections, and environment lighting; included to motivate the game-engine presentation environment, not a facility reconstruction.}
		\label{fig:fel_davinco_fidelity}
	\end{figure}

	Figure~\ref{fig:fel_paraview_zoomout} shows the full view of the FEL particle beam as rendered in ParaView from the raw point cloud, colored by velocity magnitude. The electron beam is initialized with a continuous particle distribution along all axes. As the beam traverses the undulator magnet array, the onset of microbunching becomes visible: particles self-organize into separated, periodically spaced microbunches along the direction of propagation, as revealed in the zoomed-in view in Figure~\ref{fig:fel_paraview_zoomin}. Figure~\ref{fig:fel_davinco_zoomout} shows the same data streamed through the pipeline and rendered in Unreal Engine via DaVinCo as a scene that is camera-navigable once received, using Epic's LiDAR Point Cloud plugin for point-based rendering. Figure~\ref{fig:fel_davinco_zoomin} provides a close-up of the streamed point cloud, demonstrating interactive examination of the microbunching structure with free camera navigation. Finally, Figure~\ref{fig:fel_davinco_fidelity} shows an alternate level in Unreal Engine illustrating the rendering quality a game engine offers, with physically-based reflections and environment lighting; it is included to motivate the game-engine presentation environment rather than to claim a faithful reconstruction of a facility.
	
	\subsection{Throughput and Point Scaling}
\begin{figure}[t]
		\centering
		\includegraphics[width=0.80\linewidth]{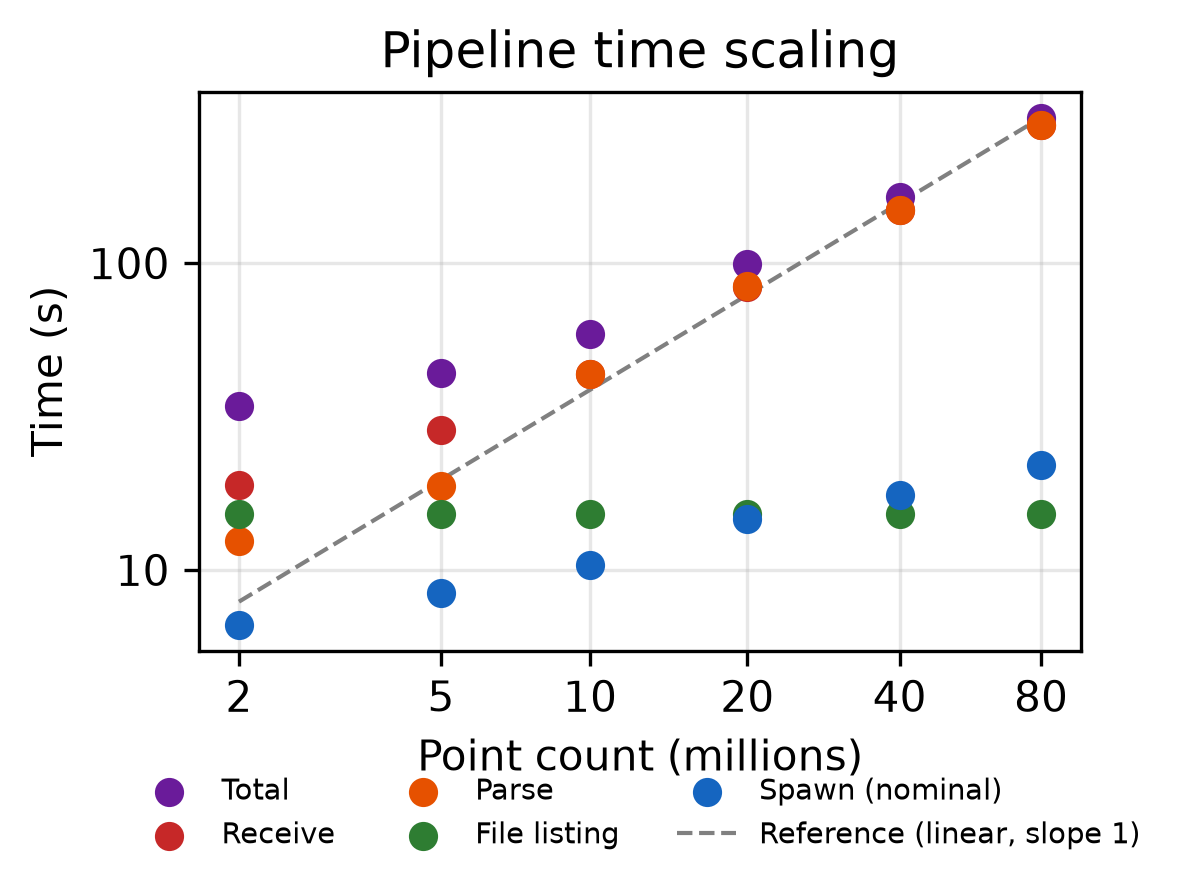}
		\caption{Payload-size scaling for the fixed 60-node deployment (log-log). Receive, parse, and total time grow near-linearly with point count, tracking the linear-scaling reference; dataset-listing time is fixed, and spawn time is shown as its nominal actor-creation trend. Node count and layout are not varied across runs, so this characterizes scaling with payload size, not with additional resources.}
		\label{fig:scaling}
	\end{figure}

	We measured the full pipeline across six dataset sizes ranging from 2 million to 80 million points on the fixed 60-node deployment of Section~\ref{sec:evaluation} (52 simulation / 8 \texttt{pvserver} nodes), varying only the payload size. Table~\ref{tab:benchmark} reports the per-run stage times across these sizes, and Figure~\ref{fig:scaling} shows the time to stream, parse, and list payloads as a function of point count for this fixed configuration. The total serialized USD payload grows from $0.12$\,GB (2 million points) to $4.77$\,GB (80 million points), and the mean per-worker-rank share grows from $1.8$\,MB to $74.6$\,MB (Table~\ref{tab:payload}). The number of \texttt{4}\,MB transfer chunks sent to the client stays nearly flat ($2121$ to $2202$) because, across this range, every per-rank USD dataset is below the \texttt{4}\,MB chunk size and is sent as a single chunk; the count therefore mirrors the point-cloud actor count, which is set by scene structure and is essentially constant across the range ($2088$--$2147$). Within this range, the $40\times$ growth in point count is absorbed as larger per-chunk size rather than more chunks. This flatness is range-specific, not structural: at the measured serialization density ($\approx$60\,MB per million points), per-rank datasets reach the \texttt{4}\,MB boundary near $150$--$160$\,M total points, beyond which oversized datasets are split into additional chunks and the count grows with total payload. The chunk size itself is configurable. All simulation points are forwarded, with no subset.
	
	Two factors dominate end-to-end time. First, the transit leg dominates the wall-clock budget: \texttt{receive} occupies the large majority of \texttt{total} time, and \texttt{parse} runs concurrently with it. Parsing operates at chunk granularity as data flows in rather than after the whole payload is assembled, which is why the reported \texttt{parse} time tracks \texttt{receive} so closely instead of exceeding it: parsing keeps pace with streaming and terminates a moment after the final chunk is processed. Second, receive and parse time scale nearly linearly with point count, paralleling the linear-scaling reference in Figure~\ref{fig:scaling}. The transit leg is request-based: throughput stays within a narrow $5.9$--$16.1$\,MB/s band across all sizes rather than tracking size, so the total time is governed by data volume and the per-chunk transit path. The band itself is bound by the round-trip latency of each $4$\,MB chunk relayed through the broker and SSH tunnel, not by the client's nominal line rate: the peak observed is well under $1$\% of the workstation's 2\,Gbit/s link. Throughput rises modestly with size as larger payloads keep more chunks in flight.
	
	The dataset-listing stage is a fixed overhead: it stays at $\approx$\,$15.2$\,s in every run regardless of size (the flat series in Figure~\ref{fig:scaling}). This is the broker's metadata round-trip to rank~0: JuSync must enumerate the per-worker dataset tables ($\sim$\,2{,}700 dataset handles in a full run) before requesting payloads, and this enumeration does not scale with payload size. Spawn time, by contrast, reflects point-cloud actor creation from the streamed data; Table~\ref{tab:benchmark} reports its nominal scaling, growing from roughly $6.6$\,s at 2 million points toward $\approx$\,$22$\,s at 80 million points. In this evaluation, no meshes or triangles are created; only point clouds are streamed.
	
	\begin{table}[htbp]
		\centering
		\caption{Per-run pipeline measurements across dataset sizes. Point counts are nominal; all simulation points are forwarded (no subset). Payload is the total serialized USD delivered to the client; times are in seconds. Receive and parse run concurrently and dataset-listing overlaps with both, so Total is the wall-clock duration, not the sum of stage columns. Spawn is the nominal point-cloud actor-creation trend (Section~\ref{sec:limitations}); other columns are measured.}
		\label{tab:benchmark}
		\small
		\renewcommand{\arraystretch}{1.0}
		\setlength{\tabcolsep}{2pt}
		\begin{tabular}{@{}lcccccc@{}}
			\toprule
			\textbf{Points} & \textbf{Total} & \textbf{Receive} & \textbf{Parse} & \textbf{Dataset List} & \textbf{Spawn} & \textbf{Payload} \\
			\textbf{(millions)} & \textbf{(s)} & \textbf{(s)} & \textbf{(s)} & \textbf{(s)} & \textbf{(s)} & \textbf{(GB)} \\
			\midrule
			2   & 34.23 & 18.92 & 12.42 & 15.23 &  6.6 & 0.12 \\
			5   & 43.99 & 28.68 & 18.75 & 15.25 &  8.4 & 0.29 \\
			10  & 58.83 & 43.52 & 43.51 & 15.25 & 10.4 & 0.59 \\
			20  & 99.29 & 83.90 & 83.94 & 15.25 & 14.6 & 1.20 \\
			40  & 164.37 & 148.83 & 148.83 & 15.26 & 17.5 & 2.40 \\
			80  & 298.17 & 282.08 & 281.49 & 15.22 & 22.0 & 4.77 \\
			\bottomrule
		\end{tabular}
	\end{table}

	\begin{table}[htbp]
		\centering
		\caption{Serialized-USD payload and transfer volume across dataset sizes (fixed 60-node deployment, Section~\ref{sec:evaluation}). Payload is the total serialized USD received by the client. \textbf{Bytes/rank} is the mean per-\texttt{pvserver}-worker-rank share (Payload\,$/\,64$ ranks). \textbf{Chunks} is the number of \texttt{4}\,MB transfer chunks sent to the client; because every per-rank USD dataset is smaller than the \texttt{4}\,MB chunk size, each dataset is sent as a single chunk, so Chunks equals the number of transferred datasets.}
		\label{tab:payload}
		\small
		\renewcommand{\arraystretch}{1.0}
		\setlength{\tabcolsep}{4pt}
		\begin{tabular}{@{}lccc@{}}
			\toprule
			\textbf{Points} & \textbf{Payload} & \textbf{Bytes/rank} & \textbf{Chunks} \\
			\textbf{(millions)} & \textbf{(GB)} & \textbf{(MB)} & \textbf{(-)} \\
			\midrule
			2   & 0.12  & 1.83  & 2121 \\
			5   & 0.29  & 4.55  & 2137 \\
			10  & 0.59  & 9.16  & 2146 \\
			20  & 1.20  & 18.75 & 2169 \\
			40  & 2.40  & 37.45 & 2184 \\
			80  & 4.77  & 74.60 & 2202 \\
			\bottomrule
		\end{tabular}
	\end{table}

	\subsection{Memory Profile}
	Host memory (\texttt{RAM}) usage rises with data volume, from $26.5$\,GB at 2 million points to $30.4$\,GB at 80 million points, and correlates at $0.79$ with point count. Because the pipeline stores serialized USD stages in per-rank memory stores on the HPC side and accumulates parsed geometry on the client, the host footprint grows roughly with the streamed data volume. The correlation is below the near-perfect values seen for time because the baseline includes the fixed Unreal Engine runtime footprint, which dilutes the data-driven portion.
	
	The modest absolute growth over a 40$\times$ range in point count ($\approx$\,4\,GB for the client side) shows that the streamed representation remains compact relative to the point count, consistent with the fixed-cost memory model of the two application hosts. GPU memory rose with the point count over the 2--10 million range ($\approx$\,$2.8$\,GB at 2 million to $\approx$\,$4.7$\,GB at 10 million) and stayed well within the RTX 5090's $32$\,GB budget; measurements at larger sizes were not reliable and are therefore omitted. Memory was not the binding constraint: the dominant resource at large scale is the transit time established above, and the runtime cost of interactivity is GPU processing of the growing point cloud, not memory pressure.
	
	\subsection{Limitations}
	\label{sec:limitations}
	
	Several aspects of the evaluation bound the current claims and motivate the future work below.
	
	\emph{Transit-bound latency.} The $5.9$--$16.1$\,MB/s throughput band is measured over the SSH-tunnel between the client Unreal instance and the cluster; it reflects the per-$4$\,MB-chunk request/response latency relayed through the broker, not the raw link or an intrinsic pipeline limit. Client-side arrival time is dominated by this serialized chunk path. Interactivity is asserted qualitatively (payloads are renderable and camera-navigable once received).
	
	\emph{Metadata round-trip.} The $\approx$\,$15.2$\,s dataset-listing stage does not scale with payload size. It reflects the broker enumerating $\sim$\,2{,}700 per-worker dataset handles at rank~0 before payload transfer, so it becomes the dominant share of end-to-end time for small payloads. The gap between datasets requested and datasets transferred in a run is not data loss: each USD dataset carries an XXH3-128 fingerprint, and datasets whose hash is unchanged from the previous timestep are skipped by the client rather than retransmitted (an intentional diff-aware deduplication). Across sizes this skipped roughly $19$--$20$\% of the $\approx$\,2{,}700 notified datasets, and every dataset that was actually transferred completed with $100$\% success and zero errors.
	
	\emph{Spawn cost reflects rebuilds, not raw sizing.} The Spawn column in Table~\ref{tab:benchmark} reports the nominal scaling of Unreal Engine's point-cloud actor creation rather than a raw per-size measurement. Raw spawn timing was non-monotonic because it follows the diff-aware deduplication: when only a few datasets change, the client updates an existing LiDAR point-cloud actor in place (cheap); when many actors must be rebuilt from new data, spawn cost rises. The reported Spawn values therefore reflect the intended Unreal actor-creation behavior at scale and should not be read as exact per-run measurements.
	
	\emph{GPU memory logging.} VRAM was only reliably captured on the $2$--$10$ million subset; larger runs produced inconsistent GPU-memory readings and are excluded. Only a bound (well within the RTX 5090's $32$\,GB) is claimed for GPU memory.
	
	\emph{No server-side or frame-time benchmark.} The cost of ANARI-USD serialization on the cluster and per-frame rendering time on the client were not instrumented; the evaluation characterizes the client arrival path (dataset-listing, receive, parse) and memory use, and no head-to-head comparison against the default disk-writing path is reported. The client and broker logs do record the total payload, per-rank byte share, and chunk counts, which are reported in Table~\ref{tab:payload}.

	\emph{Chunk count assumes a uniform split.} The \texttt{Chunks} column equals the number of transferred datasets only because every per-rank USD dataset is below the \texttt{4}\,MB chunk size, which holds for the uniform point-cloud split used here (mean per-dataset size at most $\approx$\,$2.12$\,MB at 80\,M points). Per-run logs record dataset counts but not the maximum single-dataset size, so a non-uniform split containing a dataset larger than \texttt{4}\,MB would add chunks; this is not observed for the reported data. The near-constant count is an artifact of the evaluated range: per-rank datasets cross the \texttt{4}\,MB boundary near $150$--$160$\,M total points, after which the count scales in step with payload; no dataset or payload size is structurally capped by the chunking scheme.

	\FloatBarrier
	
	\section{Conclusion}
	\label{sec:conclusion}
	
	This paper presented and evaluated an in-situ streaming pipeline that couples ParaView-based scientific processing to Unreal Engine through ANARI, USD, and ZeroMQ, exercised on a multi-node HPC deployment. By serializing USD stages directly into per-rank memory and streaming them over the ZeroMQ broker, the memory-only ANARI-USD device removes filesystem I/O from the per-timestep path, and JuSync parses the streamed geometry asynchronously and routes it to Unreal Engine's LiDAR Point Cloud plugin. Splitting producers from consumers lets clients connect or disconnect at any time without disturbing the simulation. Because the streamed payloads are expressed as USD, the path is in principle open to rendering targets beyond Unreal Engine, though only the Unreal client is demonstrated here.
	
	End-to-end evaluation on the 60-node JURECA-DC allocation: a live FEL in-situ simulation at $2$--$80$ million particles was delivered over a disk-free path from the cluster to an Unreal Engine client, with near-linear growth in receive and parse time across the transit leg, sustained throughput in a $5.9$--$16.1$\,MB/s band, and a small, bounded host memory footprint on the RTX\,5090 client. Payloads are interactable on the client once received, though arrival latency is set by the external transit link rather than by pipeline processing. The memory-only design removes the parallel filesystem from the data path, so HPC-side processing is not the scaling limit; end-to-end latency is instead governed by the external transit leg, identified as the target of future work (Section~\ref{sec:future}).
	
	\FloatBarrier
	
	\section{Future Work}
	\label{sec:future}
	
	\FloatBarrier
	
	Several directions would extend the capabilities and scalability of the current pipeline:
	
	\begin{itemize}
		\item \textbf{USD time-sampled stages}---persistent stages with time-sampled attributes reduce retransmission overhead versus discrete full payloads per timestep.
		\item \textbf{Delta-based updates}---transmit only changed geometry attributes rather than full USD payloads, improving bandwidth efficiency when timestep-to-timestep changes are small.
		\item \textbf{Volume processing}---support for USD volume data in JuSync (currently mesh and point clouds only).
		\item \textbf{Additional rendering engines}---the JuSync C API can target other USD-capable tools beyond Unreal Engine 5, such as NVIDIA Omniverse, three.js, or custom viewers.
		\item \textbf{Digital twin for OPALX}---apply the pipeline to a live in-situ digital twin of the particle accelerators at PSI using OPALX~\cite{opalx2020}. Its multi-resolution simulations stream beam and field data into a contextual Unreal replica for operational monitoring and interactive steering.
		\item \textbf{Cross-platform evaluation}---validate the pipeline on additional HPC systems and interconnect technologies beyond JURECA-DC.
		\item \textbf{Single-GPU and multi-GPU rendering boundaries}---characterize where Unreal Engine on the single RTX\,5090 client reaches its limits as the streamed point-cloud budget grows, and outline multi-GPU clients for the larger data sets motivating this work.
		\item \textbf{Forwarding-path throughput}---the measured $5.9$--$16.1$\,MB/s band is bound by the per-chunk relay over the SSH tunnel; push-based transfer, compression, a direct interconnect, or an in-cluster client would lift this ceiling. This is the precondition for the $O(10^9)$-particle regime.
		\item \textbf{Progressive / LOD streaming}---exploit the LiDAR Point Cloud plugin's dynamic level-of-detail to render a coarse representation while payloads arrive, providing early interactivity instead of waiting for the full transfer.
		\item \textbf{Simulation steering and multi-user collaboration}---steering a running simulation by adjusting parameters between payloads, and multi-user collaborative visualization over the broker's multiple simultaneous connections. The DaVinCo viewer~\cite{davinco2022} already supports a collaboration model, not yet exercised with this pipeline.
	\end{itemize}
	
	\FloatBarrier
	
	\bibliographystyle{ieeetr}
	\bibliography{references}

	\section*{Acknowledgments}
	We gratefully acknowledge the Bundesministerium f\"ur Forschung, Technologie und Raumfahrt (BMFTR) and the Ministerium f\"ur Kultur und Wissenschaft des Landes Nordrhein-Westfalen (MWK-NRW) for funding this work in the project InHPC-DE through the Gauss Centre for Supercomputing e.V. Computing resources were provided by the J\"ulich Supercomputing Centre. S.A.T.K.\ is funded by the Swiss National Science Foundation. The authors disclose that a generative AI language tool assisted with language editing of this manuscript; all scientific content, measurements, and claims are solely the authors' work.

	\section*{Reproducibility Appendix}\label{sec:appendix}
	\footnotesize
	The following notes provide sufficient information to reproduce the pipeline on a comparable setup.

	\begin{enumerate}
		\renewcommand{\theenumi}{\alph{enumi}}
		\renewcommand{\labelenumi}{(\theenumi)}
		\setlength{\itemsep}{0pt}\setlength{\parskip}{0pt}\setlength{\topsep}{2pt}

		\item \textbf{Repositories:}
		\begin{itemize}\setlength{\itemsep}{0pt}\setlength{\topsep}{2pt}
			\item \textbf{JUSYNC}: \url{https://github.com/FZJ-JSC/jusync} (middleware with C API and UE5 plugin).
			\item \textbf{ANARI-USD (ZMQ-enabled)}: \url{https://github.com/staticx-g7/ANARI-USD} (ANARI-USD with the memory-only serialization and embedded ZeroMQ transport used here).
		\end{itemize}

		\item \textbf{Build the ANARI-USD device:} Build ParaView against the ANARI SDK, then build the ANARI-USD device from the repository above; point ParaView's ANARI backend at it so the \texttt{USD} renderer is available.

		\item \textbf{Launch ParaView with the ANARI-USD device:} Start \texttt{pvserver} with MPI as needed and configure the ANARI/USD environment before launching:
		\begin{lstlisting}
export ANARI_LIBRARY='usd'
export ANARI_USD_SERIALIZE_LOCATION=$SCRATCH/$USER/anari_usd_output_pvserver
mkdir -p "$ANARI_USD_SERIALIZE_LOCATION"
export LD_LIBRARY_PATH=/p/home/jusers/$USER/jureca/parsimcat_simulation_setup/ANARI-USD/build:$LD_LIBRARY_PATH
		\end{lstlisting}
		(\texttt{ANARI\_USD\_SERIALIZE\_LOCATION} is only a working path; payloads stay in memory and are streamed, not written.)

 		\item \textbf{Start streaming from ParaView:} Load a dataset in ParaView (e.g.\ a wavelet or your simulation), enable ANARI rendering with the USD device, and apply the desired filters.

		\item \textbf{Client connection (SSH tunnel):} You must be SSH-tunneled into the machine running the broker. From the client workstation:
		\begin{lstlisting}
ssh -N -L 5556:127.0.0.1:5556 -i ~/.ssh/your_key <gateway>
		\end{lstlisting}

		\item \textbf{Launch the client (Unreal Engine or DaVinCo):}
		\begin{itemize}\setlength{\itemsep}{0pt}\setlength{\topsep}{2pt}
			\item In \textbf{Unreal Engine}, place the JuSync spawn actor in your level, configured to connect to \texttt{localhost:5556} (the tunneled broker port); starting the game connects the plugin to the broker and streamed scenes begin spawning.
			\item \textbf{DaVinCo} can be used the same way over the tunnel, with its built-in VR/desktop toolset.
		\end{itemize}

		\item \textbf{Build JUSYNC:}
		\begin{lstlisting}
cd jusync && git submodule update --init
mkdir -p build && cd build && cmake .. -DCMAKE_BUILD_TYPE=Release && cmake --build . --config Release
		\end{lstlisting}
	\end{enumerate}
	\normalsize

\end{document}